\documentclass[journal]{IEEEtran}

\usepackage{comment}
\usepackage{algorithm}
\usepackage{algorithmic}
\usepackage{cite}

\ifCLASSINFOpdf
  \usepackage[pdftex]{graphicx}
  \graphicspath{{./Figures/}{../jpeg/}}
\else
  \usepackage[dvips]{graphicx}
  \graphicspath{{./Figures/}{../jpeg/}}
  \DeclareGraphicsExtensions{.eps}
\fi
\usepackage[cmex10]{amsmath}
\ifCLASSOPTIONcompsoc
  \usepackage[caption=false,font=normalsize,labelfont=sf,textfont=sf]{subfig}
\else
  \usepackage[caption=false,font=footnotesize]{subfig}
\fi
\usepackage[hyphens]{url}

\usepackage{framed}
\usepackage{amssymb}
\usepackage{color}
\usepackage[T1]{fontenc}

\begin{document}
%

\title{Characterizing Curtailed and Uneconomic Renewable Power in the Mid-continent Independent System Operator}
%
%
%


\author{Andrew A. Chien, Fan Yang, and Chaojie Zhang \thanks{All of
    the authors are with the Department of Computer Science,
    University of Chicago, Chicago, IL 60637
    \{achien,fanyang,chaojie\}@cs.uchicago.edu, Chien is
    also with Argonne National Laboratory.}}

\maketitle

\begin{abstract}
As power grids incorporate increased renewable
generation such as wind and solar, their variability 
creates growing challenges for grid stability and
efficiency.  We study two facets: power the grid is unable to accept (curtailment), and 
power that is assigned zero economic value by the grid (negative
or zero price).  Collectively we term these {\it stranded power} or SP.

We study stranded power in the Midcontinent Independent System
Operator (MISO), characterizing quantity and temporal structure.
First, stranded power is available in the MISO grid 99\% of the time,
and often in intervals >100 hours, with characteristic seasonal and
time-of-day patterns.  Average stranded power often exceeds 1 GW, with
duty factors as high as 30\%.  About 30\% of all wind generation in MISO
is stranded.  Examination of the top 10 individual
sites shows stranded power can be as high as 70\% duty factor and
250MW.  Trends over the past 3.5 years suggest stranded power is a
persistent phenomenon.  The study characterizes opportunities to
exploit stranded power.

We consider using energy storage to increase the utility of
stranded power.  For a range of power levels and uniformly-distributed
storage, adding 5 hours of storage doubles
duty factor to 30\% at 4MW, but another 95 hours
is  required for the next 15\% increase.  At 
4MW with 50 hours of storage, only 3 of 200 sites reach
100\% duty factor, and with 100 hours required for the next 10 sites.  
Higher power levels require 100's of hours.
Storage at the top 10 sites is more productive, 5
hours increases duty factor to 70\% at
4MW, but further storage has diminishing benefits.  
Studies of the
amount of power served by storage show that distribution to
the best sites provides 2 to 3.7-fold advantages over uniform
distribution.

\end{abstract}


%
\IEEEpeerreviewmaketitle



\section{Introduction (1 page)}
\label{sec:introduction}

Over the past two decades, a growing consensus 
on climate change due to 
anthropogenic carbon has emerged
\cite{IPCC-CC2014,Gore06}.  In response, there are growing
worldwide efforts to
reduce the amount of carbon being released into the atmosphere
\cite{Kyoto1997,Paris2015}. 

Ambitious ``renewable portfolio
standards'' (RPS) goals for renewable power as a fraction of overall
power have been widely adopted.  Midwest examples from the
Mid-continent Independent System Operator (MISO) system include Illinois (25\% by 2025) 
to Minnesota ($25\sim31$\% by 2025).  California and New York
have adopted a 50\% goal by 2030 \cite{rps-california,Megerian15}. 
Obama's ``Clean Power Plan,'' (August 2015) \cite{ObamaCPP15}, calls for a 
32\% national reduction in electric power carbon emissions by 2030, with renewables a
critical element.  The U.S. Department of Energy's
landmark report, ``Wind Vision 2015,'' targets a United States 35\% RPS for wind alone by 2050,
with regions such as the Midwest and Texas  at 50\% wind RPS.
A recent report, proposes
50\% solar-only RPS scenario for California \cite{Storage-California-50RPS-2016}.  
Any of these
scenarios are
a dramatic jump from the US's combined
solar and wind RPS of 5.2\% in 2014 \cite{renewable2014}.  
These ambitious and transformative goals pose serious power grid challenges 
including ability to achieve
``merit order'', efficiency, stability, and resiliency. 

Evidence of such challenges include growing curtailment, uneconomic
generation, and RPS stagnation.  Curtailment, where the power grid is unable to accept
renewable generation due to congestion, misforecast, or excess
generation, causes power to be discarded at the generation site
(Europe and the United States \cite{Lew2013,Bird2013} and China
\cite{GWEC-Annual16}).  Despite programs to increase transmission
capacity and employ economic dispatch, curtailed power in Europe,
United states, and China exceeds 50 TWH per year \cite{GWEC-Annual16}.
Economic dispatch has been deployed to reduce curtailment, providing
economic incentives (payments) and disincentives (negative payments)
for generation, but the result is uneconomic power generation, power
purchased by the power grid at a negative price
\cite{ERCOT-negative-price15,Germany-negative-price16}.  We use the
term {\it stranded power} (SP) to describe both curtailed power and
uneconomic generation as they together represent excess renewable
power generation.
As several regional power grids have reached RPS approaching and in
some cases exceeding 30\%, there is growing evidence that achieving
RPS of 50\% poses significant challenges in grid flexibility
(RPS growth stalled \cite{GermanRPS-stop2016}) growing stranded power 
\cite{CurtailmentUS14,Solar-Challenges16,E3report,Texas-California-too-much2016,ERCOT-negative-price15,ERCOT-free-power15,Germany-negative-price16}, and grid reliability challenges
\cite{MISO-Revenue-Sufficiency2005,Solar-Challenges16}.  

All of these scenarios suggest that future grids will have much larger
quantities of stranded power than todays.  


Thus, to provide a basis for understanding and exploitation of stranded power, 
we undertake a detailed analysis of the dynamic properties
of stranded power.  Our objective is to provide insights into its
current properties that might enable its exploitation for use or
inspire new techniques that reduce its occurrence.  To this end, we
analyze 40 months of detailed records from the
Midcontinent Independent System Operator (MISO) power grid market.  These
records include detailed temporal structure (5-minute intervals), potential
generation (eco\_max), actual accepted generation, and pricing.  Based on 
analysis of this data, our specific contributions include:

\begin{itemize}
\item A detailed temporal characterization of stranded power in the
  MISO grid, showing it is available 99\% of the time, and often in
  intervals with long duration (mean of 109 hours, stdev of
  110 hour, and max of 816 hours).  Significant quantity is
  is available, averaging over 1 GW and with duty factor 30\% at
  that level.  Significant daily (hours) and seasonal variation are documented.

\item Study of a 3.5-year period that demonstrates the persistence
of stranded power in the MISO grid of more than 6 TWh and up to 15 TWh per year

\item Considering stranded power locally, we find single-site stranded power
  can have duty factors as high as 70\% and power levels as high as 250 MW.  
  hours) enables a more usable duty factor of 40\% drawn from intervals >5
  hours.  Those with highest quantity of stranded power 
  can achieve duty factors >30\% with
  intervals >1 hour and at significant power.  Studies of top 10 sites demonstrate 
  similar characteristics.


\item Increasing stranded power usability with energy storage requires
  defining storage capacity and power as well as stranded power load.
  At 0.25GW load, the first hour of storage increases duty factor to 82\% (a
  15\% increase), but benefits decrease rapidly,
  with minimal benefits beyond 8 hours.  At higher power levels, the
  achievable duty factor falls off significantly with 62\% and 46\%
  achievable for 1 GW and 2 GW respectively.

\item Adding storage  at carefully selected sites is productive.  
  At the best 5 sites, adding 5 hours increases duty factor
  over 70\% at 4 MW.  Selective distribution to the best sites
  achieves 2 to 3.7x greater benefit than uniform distribution.
  However, attaining 100\% duty is difficult.  At a 4 MW power level,
  50 hours is sufficient for only 3 wind sites to reach 100\% duty, and
  250-1000 hours required at higher power levels.  


\end{itemize}

The rest of the paper is organized as follows.  In Section
\ref{section:background} we briefly summarize realities of the modern
power grid - power markets, grid dispatch, and the incorporation of variable
renewable generation.
Section \ref{section:characterization} we define stranded power, and characterize
it within the MISO grid.  A common proposal is to use energy storage to increase
the utility of stranded power; we explore the rewards for that approach in
Section \ref{section:enhanced}.  We discuss our results in the context
of 
related work in Section \ref{section:related}.  Finally, we summarize 
our results and point out several promising directions for future research in Section \ref{section:summary}.

\section{Background}
\label{sec:background}
\label{section:background}

We briefly summarize key background, including how modern power grids dispatch generation,
the global push to renewal-based power generation, and
the Midcontinent Independent System Operator, a power grid 
that we study in detail in this paper.

\subsection{Modern Power Grid Dispatch}

Power grid management is difficult 
because the grid must
match producers and loads instantaneously -- power is not
stored in significant quantities.  Furthermore, the power
grid must accommodate sudden increases increases or decreases
in load due to weather, equipment failures, or sunrise.  If they
cannot, power outages occur \cite{San-diego-outage}.
Advanced research and technology is being pursued for
energy storage and power switching, but such technologies are
not a significant factor in today's power grids. 
Thus, if power cannot be productively transmitted to a load, then 
it is wasted.  

While energy markets vary, modern ISO's in the US dispatch 
generation and price power purchases based on a fast-moving
dynamic market system.  California's ISO uses a market that prices
and dispatches power in 
12-minute intervals, and the Midcontinent ISO (MISO) market that we
study is 
even faster, using 5-minute dispatch and pricing intervals.  These
real-time markets set transaction prices for megawatt-hours (MWh) 
of power, determining the prices paid by utilities for power
generators, by the grid to generators, as well as charges for 
transmission.

Because transmission is limited, 
ISO power markets generate 
locational marginal pricing (LMP), that is a price for electricity
at each distinct node in the transmission network.  These nodes typically
include generation sites, intermediate
nodes, and 
egress to utility distribution.  The markets algorithms accept offered
generation and pricing ``bids'', for an array of power levels, are designed
to be fair to different market participants, and also achieve desirable
objectives such as ``merit order'', purchase first from generators of lowest cost, and
priority to renewables, minimize carbon-based genearation, and so on.  
However, due to ramp and transmission congestion constraints, 
the markets cannot do so perfectly. In practice,  
the LMP varies widely by location at a particular time
as well as at a given location at various points in time.  
Even in adjacent 5-minute intervals, prices can swing by \$100 or more \cite{caiso-lmp}.  
While the dynamic range
varies by ISO, in the MISO grid, prices can vary from +\$1,000 to -\$1,000 per MWh.
In 2014, the average wholesale price of power in MISO was \$30/MWh.





\subsection{Growing Renewable Power and RPS Goals}

Growing concerns about carbon emissions and its long-term impacts on
climate change have created a world-wide consensus to
increase renewable-based power generation.  Notably, wind and
solar generation, are the most rapidly growing sources, and
both been the subject of numerous
government programs to encourage their deployment and use, including
``feed-in'' tariffs in Germany, Spain, and other nations in Europe, as
well as ``production tax credits'' in the United States.
In the US, solar and wind generation together
comprised 5.2\% of overall power in 2014
\cite{renewable2014}.  California has been a leader for 
setting Renewable Portfolio Standards (RPS), requirements for 
power generation mix, reaching a 20\% renewable mix in 2010, and 
on track to reach its 33\% target for
2020 \cite{rps-california} for wind and solar power.  In September 2015, California adopted
an RPS goal of 50\% renewable by 2030 \cite{California-50RPS}. 
Other states across the midwest (included in the MISO power
grid) have adopted a range of standards ranging from 25\% (2015) in Illinois, 25-31\% (2025)
in Minnesota, and 55\% (2017) in Vermont.  Other large states include
50\% by 2030 in New York, and 10GW by 2025 in Texas.


Obama's ``Clean Power Plan'' (August 2015) \cite{ObamaCPP15}, calls for a 
32\% national reduction in electric power carbon emissions by 2030, with renewables a
critical element.  The U.S. Department of Energy's
landmark report, ``Wind Vision 2015,'' targets a United States 35\% RPS for wind alone by 2050,
with regions such as the Midwest and Texas  at 50\% wind RPS.
A recent report, proposes
50\% solar-only RPS scenario for California \cite{Storage-California-50RPS-2016}.  
Europe has been the most aggressive in deploying renewables \cite{Bird2013}, and while 
starting later, China has rapidly grown its wind generation, in 2015 becoming the world's 
largest wind generator with over 145 GW of installed wind generation \cite{GWEC-Annual16}.

Renewables such as wind and solar have time-varying productivity.
Solar follows a diurnal cycle, but has significant variation within
that.  Wind power has higher variability over long periods (weeks or
months), but much less over short periods (hours).  Both wind and
solar, particularly in the US, are deployed in distributed fashion
which when combined with variation creates major power grid scheduling
and transmission challenges, giving
rise to the phenomenon of stranded power \cite{Lew2013,Bird2013}, which
is widely viewed as growing in magnitude with RPS \cite{WindVision}.

\subsection{Midcontinent Independent Independent System Operator}
\label{section:miso-background}

The Mid-continent Independent System Operator (MISO) is 
is one of the largest power markets in the United
States, and is the focus of our study.  MISO shares its market data
openly.  MISO manages power
for a large geographic area, that covers most of ten states (Illinois,
Indiana, Iowa, Minnesota, Wisconsin, Michigan, North Dakota, Arkansas,
Mississippi, Louisiana, as well as parts of Texas, Montana, Missouri,
and Manitoba.  MISO serves over 42 million people, and in 2014, transacted
\$37 billion of power, and assessed \$2.2 billion in power
transmission charges.  It governs more than 65,000 miles of transmission
lines, has 2,000 pricing nodes\footnote{MISO sets prices for these nodes every 5 minutes.},
and over 400 market participants.  For all of these, it sets prices every 5 minutes.
MISO's historic peak load is 130 GW,
and it provides over 500 TWH on an annual basis.  For more information,
see http://www.miso.com/.

\subsection{Potential Uses of Stranded Power}

An important motivation for characterizing stranded power is to enable
its profitable exploitation.  While exhaustive enumeration is
infeasible, economically viable and effective use is constrained by
several key factors: low capital equipment cost, interruption and
delay-tolerance, location-insensitive, and of course energy-intensive
with high-value output.  Equipment cost and interruption-tolerance are
critical, as lower duty factors than reliable grid power will be
achieved.  Location-insensitive, as the greatest quantities of
stranded power occur in remote regions.  Naturally, power-intensive,
high-value outputs give greatest economic advantage.  Potential examples
include cloud computing \cite{chien2015zero,YangChien15}, water-splitting to make
hydrogen fuel \cite{Water-splitting}, fertilizer
manufacture \cite{Haber-process},  bitcoin
mining \cite{AM-S9-2015}, and even lightweight manufacturing \cite{Digital-Manufacturing2015}.  Other 
environmentally relevant tasks include water desalination \cite{SanDiego-desalination2015}
and even carbon-scrubbing \cite{CO2-tower14}.

\section{Characterization of Stranded Power}
\label{sec:characterization}
\label{section:characterization}

We analyze power grid behavior to understand properties of stranded
power, including quantity, as well as temporal and spatial distribution.  


\subsection{What is Stranded Power?}

We use the term {\it stranded power} to describe both curtailed power
and uneconomic generation as they together represent excess renewable
power generation.  Stranded power arises from variability in renewable
generation, combined with constraints in grid management such as
ramps, transmission limits, and merit-order.  These complex,
interacting constraints must be solved for each time interval,
5-minutes in the MISO grid, for example.  When the resulting schedule
cannot accomodate power from a generator, {\it curtailment} results, and the
generator's power is excluded from the grid.  With deregulation, modern
power grids use market-based dispatch, generally setting prices by a
publicly-declared algorithm called locational marginal pricing (LMP)
\cite{LMP}.  Thus, unneeded power generation is discouraged by market
signals (prices), that fluctuate over wide ranges in periods as short
as 5 minutes.  In these markets, negative pricing is quite common, due
to a combination of zero fuel cost and externalities (e.g. production
tax credits).  When the market assigns power a negative price, we term 
that power {\it uneconomic}, as though the power is accepted into the grid, the generator
literally pays the grid to accept it.


\begin{figure}[tbh]
\centering
\vspace{-0.1in}
\includegraphics[width=3.4in]{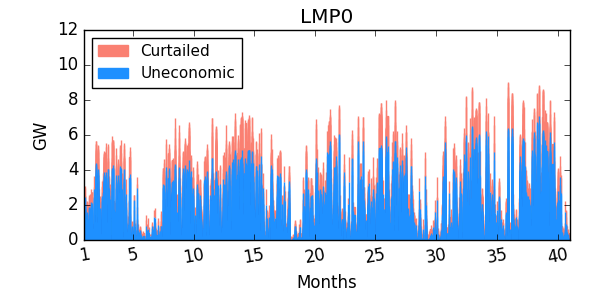}
\vspace{-0.1in}
\caption{Stranded wind power in the MISO grid, Curtailed (red) and Uneconomic (blue), March 1, 2013-June 30, 2016. (LMP0)}
\vspace{-0.08in}
\label{fig:grid_quantity_vs_time_classify_lmp0}
\label{figure:stranded_power_lmp0}
\end{figure}

Historical stranded power for the MISO grid is show in Figure
\ref{figure:stranded_power_lmp0} for a 3+ year period.  The level of
stranded wind power grid-wide regularly reaches 1 to 2 GW, and varies
dramatically, exceeding 4GW regularly, and occasionally reaching 7 GW.
While curtailed power (red) is significant, the uneconomic power
component accounts for the largest part of stranded power.  The LMP0
model counts the power delivered at negative price as uneconomic power
(see Section \ref{section:stranded-power-models}).  Thus, stranded
power is frequent and widespread
\cite{ERCOT-negative-price15,Texas-California-too-much2016,Germany-negative-price16,MISO};
and can be characterized precisely by analysis of power markets.
Stranded power is both a significant challenge for
renewable generator economics, and represents the current ability of the power
grid to integrate variable renewable generation.  The latter is an important 
indication of future challege as grids move to higher renewable fractions (renewabe-portfolio standar or RPS).


\begin{table}[tb]
\renewcommand{\arraystretch}{1.3}
\vspace{-0.1in}
\caption{MISO Market Data (Real-time Offers (RTO))}
\label{tab:miso-data-set}
\label{tab:rto}
\vspace{-0.1in}
\centering
\begin{tabular}{l|l}
\hline
Parameter & Value \\ \hline
\hline
Period  & 3/1/2013\textemdash 7/1/2016 \\ \hline
Generation Sites & 1,329 Total, 206 Wind \\ \hline
5-minute Intervals & 207,621,612 Total, 59,459,126 Wind \\ \hline
Total TWh & 1951.58 Total, 136.63 Wind \\ \hline
Total \$'s Power & \$56.8 B Total, \$2.4 B Wind \\ \hline

\end{tabular}


\vspace*{0.05in}
\begin{tabular}{l|l}
\hline
Real-Time Offer (value) & Description\\
\hline \hline
Time & Start Time of the 5-minute interval \\
\hline
Economic Max & Power offered by generator (next interval)\\
\hline
Delivered MW & Delivered power for the interval\\
\hline
LMP & Locational Marginal Price for interval \\
\hline
\end{tabular}
\vspace{-0.08in}
\end{table}


\subsection{Stranded Power in the Midcontinent USA}

To empirically characterize stranded power, we analyze the
real-time market cleared offers (RTOs) for the Midcontinent
Independent System Operator (MISO) \cite{MISO} power market as
described in Table \ref{tab:miso-data-set}.  More detail on the MISO power grid
is given in Section \ref{section:miso-background}.  The RTO's include
locational marginal price (LMP), offered power, cleared power, offered
price, and a wealth of other data for 5-minute intervals.  
MISO includes significant generation from coal, nuclear, and natural gas, but we focus
on the
largest source of renewable power, wind turbines that account for $\approx$
10\% of MISO's power.  Thus, the following analysis focuses on MISO 
wind generation sites exclusively.  
Using the RTO's, we compute curtailed power
as the difference between \textit{Economic Max} and \textit{Delivered
  MW}.\footnote{Some smaller sites do not bid
\textit{Economic Max}, allowing MISO to forecast for
them.  This service was offered by MISO to ease integration
for small generators, and produces RTO's with no
\textit{Economic Max} values.  
As a result, these sites account for only 9.6\% of the wind power.  We
exclude these sites, reducing the sites consider from 206 to 137 and
thus our measurements underestimate stranded power.}

\begin{figure}[htb]
\centering
\vspace{-0.1in}
\includegraphics[width=0.75\columnwidth]{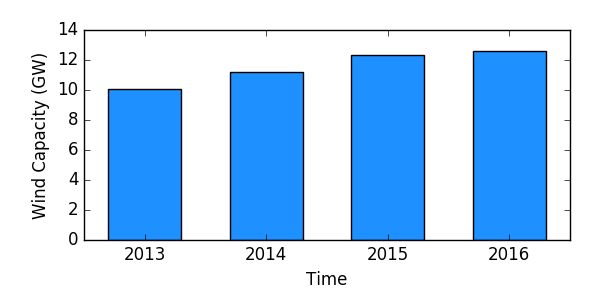}
\vspace{-0.1in}
\caption{MISO Grid wind nameplate capacity}
\vspace{-0.08in}
\label{fig:grid_yearly_capacity}
\end{figure}

\begin{figure}[htb]
\centering
\vspace{-0.1in}
\includegraphics[width=0.75\columnwidth]{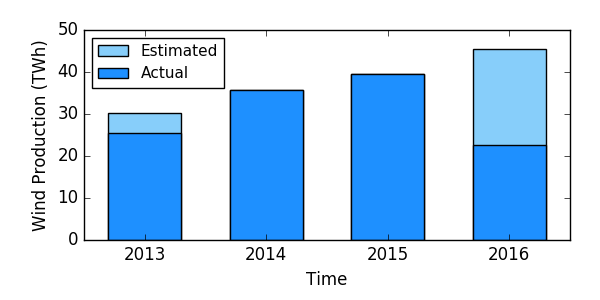}
\vspace{-0.1in}
\caption{Wind generation in MISO.  Estimates for 2013 due to market
startup January-February, and for 2nd half 2016.  The estimates are 
  scaled to 12 months based on seasonal variation.}
\vspace{-0.08in}
\label{fig:grid_yearly_production}
\end{figure}

We document the wind generation nameplate capacity for the MISO grid
(see Figure \ref{fig:grid_yearly_capacity}) that has grown steadily,
from a base of 13 GW in 2013 to nearly 15 GW in 2016.  Wind production
has grown even faster, approaching 40 terawatt-hours (TWH) in 2015,
and in a seasonally adjusted projection expected to exceed 40 TWH in
2016 (see Figure \ref{fig:grid_yearly_production}).  Analysis of key
statistics for wind generation sites shows a decided shift to larger
wind sites (see Table \ref{tab:generator_capacity_by_year}), the
largest site is nearly 500 MW.


\begin{table}[htb]
\vspace{-0.1in}
\caption{MISO Wind Generator Capacity (MW)}
\label{tab:generator_capacity_by_year}
\vspace{-0.1in}
\centering
\begin{tabular}{l|r|r|r|r|r}
\hline
Year & Total & Sites & Mean & StdDev & Max \\
\hline \hline
2013 & 10,048 & 115 & 87 & 62 & 289 \\
\hline
2014 & 11,194 & 128 & 87 & 64 & 356 \\
\hline
2015 & 12,312 & 135 & 91 & 69 & 497 \\
\hline
2016 & 12,590 & 137 & 92 & 69 & 497  \\
\hline

\end{tabular}
\vspace{-0.08in}
\end{table}

\subsubsection{Stranded Power Models and Metrics}
\label{section:stranded-power-models}


We define two families of stranded power models.  First, instantaneous
stranded power is defined as the power delivered in the 5-minute
intervals where the locational marginal price is negative or zero; we
call these models LMP(C), and consider LMP0.  Second, because while power is transacted
in 5-minute intervals, power is generally only useful in intervals of
hours, if not days, we define the net price model, NP(C), where the
average power price over a set of contiguous 5-minute intervals is C
dollars/MWh, considering NP0 and NP5.  We define each of these
formally below:

\noindent\textbf{Instantaneous Stranded Power:}  LMP(C)
\begin{equation}
LMP<C  ~~~  where ~~ C = price~ threshold
\end{equation}

\noindent\textbf{Net Price Stranded Power:} NP(C)
\begin{equation}
NetPrice<C  ~~~  where ~~ C = price~ threshold 
\end{equation}
\begin{equation}
NetPrice = \frac{\sum_{period}LMP\cdot Power}{\sum_{period}Power}, Power\textrm{~in MWh}
\end{equation}

We apply the LMP0, NP0, and NP5 stranded power models to the MISO
market records.  LMP0 represents the corresponds to the simplest
defintion of uneconomic power.  NP0 represents a more flexible
definition of uneconomic, and NP5 represents power that may in fact be
uneconomic for wind generators (absent subsidies).  To characterize
stranded power we use the following metrics: instantaneous {\it quantity}
(MW or GW), {\it aggregate quantity} (GWh or TWh), and {\it duty factor}
(fraction of time available).  An important characteristic for
usability are the periods of contiguous time that stranded power is
available, an \textit{interval}.  We study {\it distributions} of stranded
power interval durations.


\begin{figure}[tbh]
\centering
\vspace{-0.1in}
\includegraphics[width=3.4in]{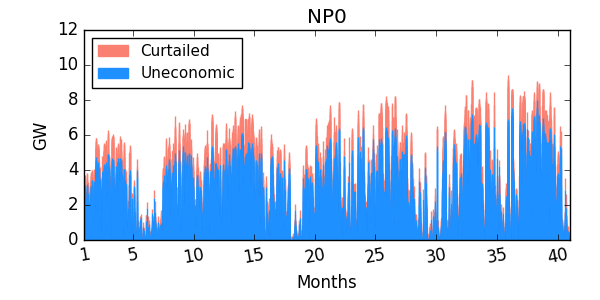}
\vspace{-0.1in}
\caption{Curtailed (red) and Uneconomic (blue) wind generated power in the MISO grid, March 1, 2013-June 30, 2016.  (NP0).}
\vspace{-0.08in}
\label{fig:grid_quantity_vs_time_classify_np0}
\end{figure}

\begin{figure}[tbh]
\centering
\vspace{-0.1in}
\includegraphics[width=3.4in]{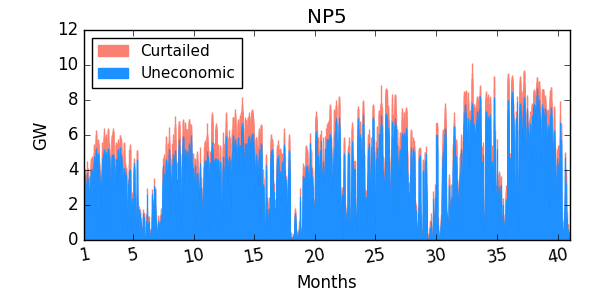}
\vspace{-0.1in}
\caption{Curtailed (red) and Uneconomic (blue) wind generated power in the MISO grid, March 1, 2013-June 30, 2016.  (NP5).}
\vspace{-0.08in}
\label{fig:grid_quantity_vs_time_classify_np5}
\end{figure}

\subsection{Stranded Power for the Entire MISO Grid}


We first explore the temporal behavior of stranded power, looking across
the entire grid.  This broad view provides a global characterization
of the aggregate quantities of a rapidly fluctuating, distributed
phenomenon.  This global characterization shows both the limitations
of today's grid in absorbing wind generation, and the upper bound on
what stranded power could be captured and exploited.  First, Figure
\ref{fig:grid_quantity_vs_time_classify_lmp0}, already discussed,
shows instantaneous stranded power, LMP0, peaking at 8 GW, has steep
fluctuations, and shows uneconomic power as much larger than
curtailment.  Using a more flexible definition, NP0, not only
increases the quantity of stranded power significantly, but also
smooths the troughs significantly (see Figure
\ref{fig:grid_quantity_vs_time_classify_np0}).  A critical challenge
for renewables is economic viability, so NP5 allows a small amount of
money to be paid for power (\$5/MWH, about one-fifth the market
price).  NP5 increases the quantity of stranded power signficantly and
further smooths the troughs (see Figure
\ref{fig:grid_quantity_vs_time_classify_np5}).

\begin{figure}[htb]
\centering
\vspace{-0.1in}
\includegraphics[width=3.4in]{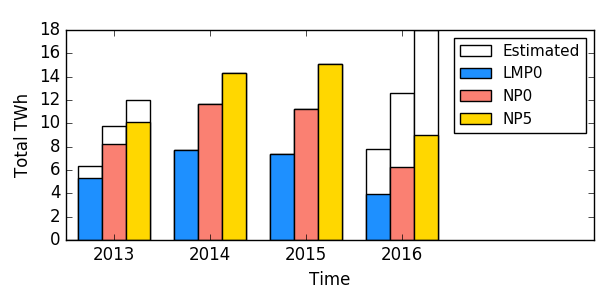}
\vspace{-0.1in}
\caption{Total stranded power for the MISO Grid (2013-2016), Various models.}
\vspace{-0.08in}
\label{fig:grid_yearly_sp}
\end{figure}

Looking at the three stranded power models together (Figure
\ref{fig:grid_yearly_sp}), we can see that shifting from an
instantaneous pricing model to an average price model, NP0, increases the
quantity of stranded power  by nearly 30\%.  In NP5, adding the
\$5/MWH increases the available stranded power by another 15-20\%.
Comparing the stranded power totals to 
the total wind production in
Figure \ref{fig:grid_yearly_production}, shows that a large fraction, approximately
30\%, of MISO's wind power is stranded.  


\subsection{Temporal Properties: Duty, Duration, Correlation}

\begin{figure}[htb]
\centering
\vspace{-0.1in}
\includegraphics[width=0.75\columnwidth]{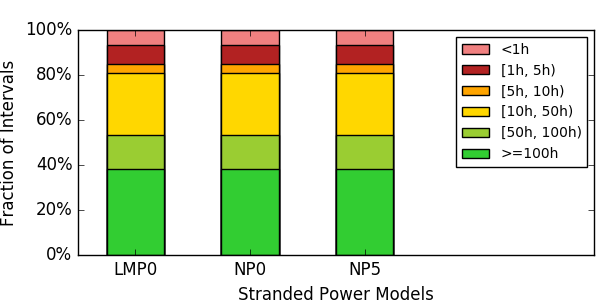}
\vspace{-0.1in}
\caption{Interval statistics for the entire grid (average by count of intervals).}
\vspace{-0.08in}
\label{fig:grid_interval_statistics_count}
\end{figure}

\begin{figure}[htb]
\centering
\vspace{-0.1in}
\includegraphics[width=0.75\columnwidth]{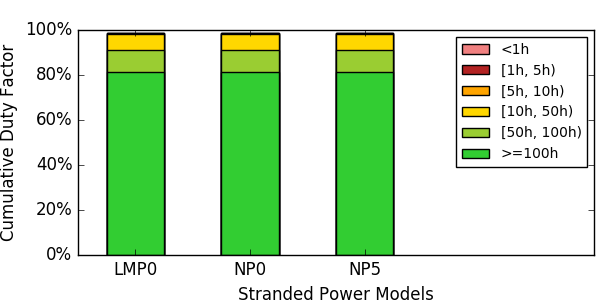}
\vspace{-0.1in}
\caption{Interval statistics for the entire grid (contribution to duty factor).}
\vspace{-0.08in}
\label{fig:grid_interval_statistics_duty}
\end{figure}

The key challenge for stranded power utilization is its intermittence.
Reliability in power grids is a critical feature, though outages are
inevitable.  To evaluate the usability of stranded power we consider
the time intervals that it is available and their duration.  In Figure
\ref{fig:grid_interval_statistics_count} we show the fraction of
intervals of each duration, by count; surprisingly there are many
stranded power intervals longer than 10, 50, even 100 hours.  However,
weighting by count overweights short intervals that are numerous, but
contribute little to temporal availability.
To understand how intervals of various durations
contribute to duty factor, we plot this directly 
in Figure \ref{fig:grid_interval_statistics_duty}.
The resulting plot clearly shows that the vast majority of duty factor
comes from long intervals, more than 50 or even 100 hours long.  Further,
the duty factor for stranded power, grid-wide, is 99.8\%; there is stranded
power is present in the grid nearly all of the time.
Interval statistics for all three models are presented in
Table \ref{tab:grid_interval_statistics}, and show that the average interval
duration in longer than 100 hours, and the longest interval is over 800 hours (more
than a month).

\begin{table}[htb]
\vspace{-0.1in}
\caption{Statistics for Stranded Power Interval Lengths, entire grid (Hours) by Model}
\label{tab:grid_interval_statistics}
\vspace{-0.1in}
\centering
\begin{tabular}{l|r|r|r|r|r}
\hline
Year & \#Intervals & Total & Mean & StdDev & Max \\
\hline \hline
LMP0 & 263 & 28,903 & 109.9 & 131.3 & 816.5 \\
\hline
NP0 & 263 & 28,903 & 109.9 & 131.3 & 816.5 \\
\hline
NP5 & 263 & 28,903 & 109.9 & 131.3 & 816.5 \\
\hline
\end{tabular}
\vspace{-0.08in}
\end{table}

\begin{figure}[htb]
\centering
\vspace{-0.1in}
\includegraphics[width=3.4in]{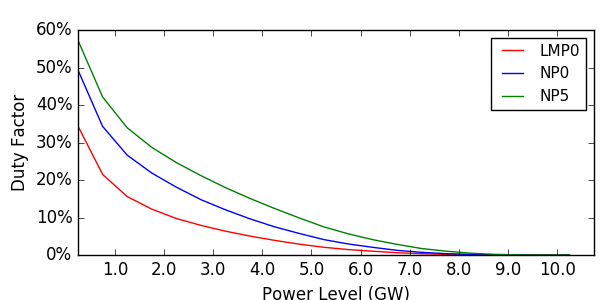}
\vspace{-0.1in}
\caption{Duty factor vs. power level for whole grid.}
\vspace{-0.08in}
\label{fig:grid_duty_vs_powerlevel}
\end{figure}

Next, we consider the duty factor that can be achieved at various
power levels, to characterize how much power is available.  As the power level increases,
more intervals will be characterized as ``insufficient'', and switch from stranded power
available to an outage.  As shown in Figure \ref{fig:grid_duty_vs_powerlevel}, the duty factors
begin at 55\%, 48\%, and 33\% for NP5, NP0, and LMP0 respectively, and decrease as the power
requirement is increased.  At 1 GW, the duty factor range from 40\% down to 20\%.


\begin{figure}[htb]
\centering
\vspace{-0.1in}
\includegraphics[height=1.25in,width=0.5\columnwidth]{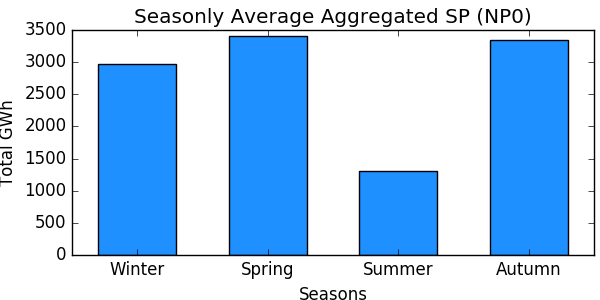}
\vspace{-0.1in}
\caption{Total stranded power, Seasonal average for the entire
  grid. March 21, 2013 - March 20, 2016 (three years).}
\vspace{-0.08in}
\label{fig:grid_seasons}
\end{figure}

Societal activity often has a diurnal, weekly, or even seasonal
structure; we explore how stranded power varies with these temporal
periods.  Beginning with seasons, Figure \ref{fig:grid_seasons} shows
the aggregate stranded power by season, showing that wind stranded
power correlates with high levels of productivity in fall and spring,
but is anticorrelated with load which peaks in the summer for MISO.


\begin{figure}[htb]
\centering
\vspace{-0.05in}
\includegraphics[width=0.8\columnwidth]{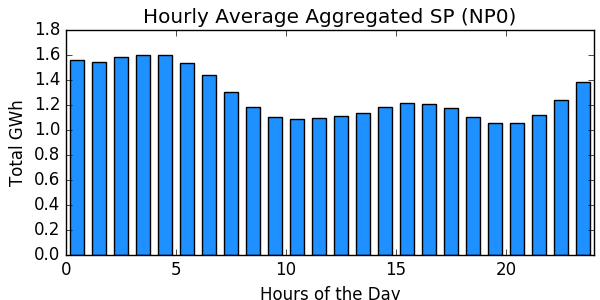}
\vspace{-0.1in}
\caption{Hourly average stranded power for the entire grid.}
\vspace{-0.08in}
\label{fig:grid_hours}
\end{figure}

Next we consider time of day, plotting the stranded power against time
of day (see Figure \ref{fig:grid_hours}), and using the more than one
thousand days in the period of study.  The diurnal trends are clear,
with 100's of MW more stranded power available on average during
late-night, early morning hours (off-peak).  While stranded averages
around 1.5 GW, the peaks exceed 8 GW (see Figure
\ref{fig:grid_quantity_vs_time_classify_np0}).

\begin{figure}[htb]
\centering
\vspace{-0.1in}
\includegraphics[width=0.8\columnwidth]{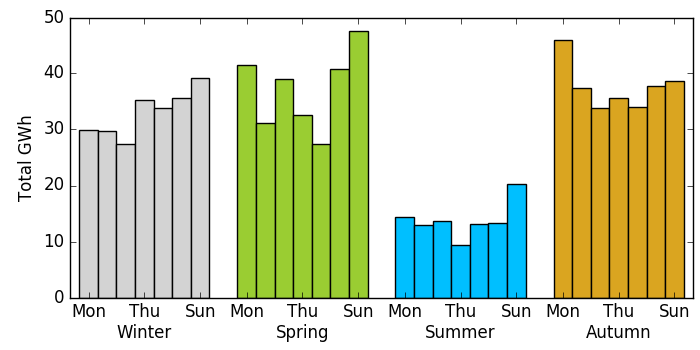}
\vspace{-0.1in}
\caption{Combined seasonal and day of week averages for entire grid. March 21, 2013 - March 20, 2016 (three cycles of seasons).}
\vspace{-0.08in}
\label{fig:grid_season_weekdays}
\end{figure}

\begin{figure}[htb]
\centering
\vspace{-0.1in}
\includegraphics[width=\columnwidth]{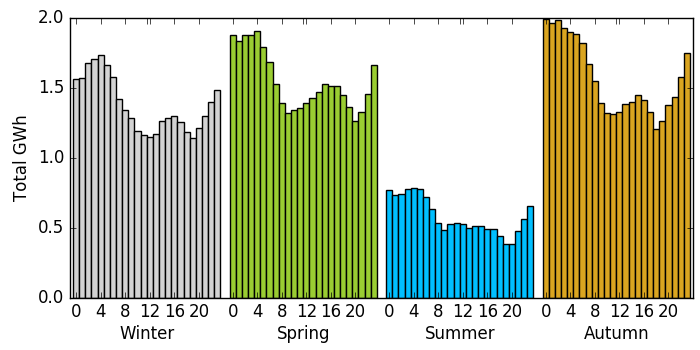}
\vspace{-0.1in}
\caption{Combined seasonal and hourly averages for entire grid. March 21, 2013 - March 20, 2016 (three cycles of seasons).}
\vspace{-0.08in}
\label{fig:grid_season_hours}
\end{figure}

The seasonal variation of stranded power is so striking that we also
combined it with day-of-week (see Figure
\ref{fig:grid_season_weekdays}) and hourly binning (see Figure
\ref{fig:grid_season_hours}) to determine if the different temporal structures are independent.
The figures suggest that they are largely independent, showing striking seasonal variation 
in magnitude of average stranded power in both graphs.  The weekly substructure seems less 
uniform across seasons, but the hourly substructure is quite uniform.

\section{Local Stranded Power (Individual Wind Sites)}

\begin{figure}[htb]
\centering
\vspace{-0.1in}
\includegraphics[width=0.8\columnwidth]{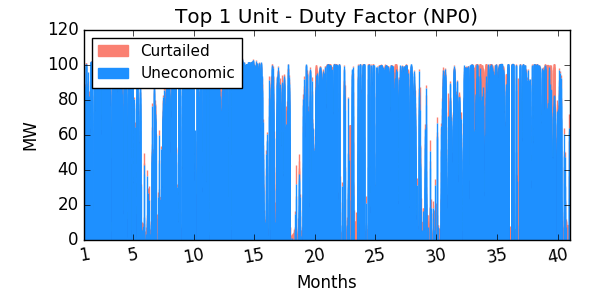}
\vspace{-0.1in}
\caption{Stranded power Quantity vs. time for site with highest duty factor.}
\vspace{-0.08in}
\label{fig:top1duty_quantity_vs_time_np0}
\label{figure:top-duty-np0}
\end{figure}
\begin{figure}[htb]
\centering
\vspace{-0.1in}
\includegraphics[width=0.8\columnwidth]{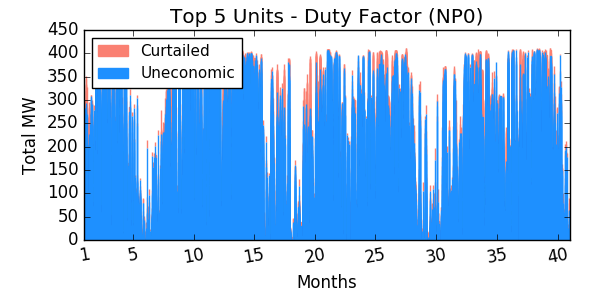}
\vspace{-0.1in}
\caption{Stranded power Quantity vs. time for 5 sites with highest duty factor.}
\vspace{-0.08in}
\label{fig:top5duty_quantity_vs_time_np0}
\label{figure:top5duty-np0}
\end{figure}

Next, we consider the properties of stranded power at single sites.  A
local view of stranded power not only characterizes the worst case for
a wind generator -- how much power is being stranded -- and the
potential for exploitation -- via colocation.  Here we consider the
most extreme stranded power sites by three criteria: 1) the top duty
factor, 2) the top 5 duty factor, and 3) the top 5 by quantity.
First, looking at the top duty factor site, Figure
\ref{figure:top-duty-np0} shows that this site has 100 MW of stranded
power in many intervals.  Looking across the top 5 duty factor sites,
stranded power can be a high as 400 MW, with large quantities (see
Figure \ref{fig:top5duty_quantity_vs_time_np0}).  

\begin{figure}[htb]
\centering
\vspace{-0.1in}
\includegraphics[width=0.8\columnwidth]{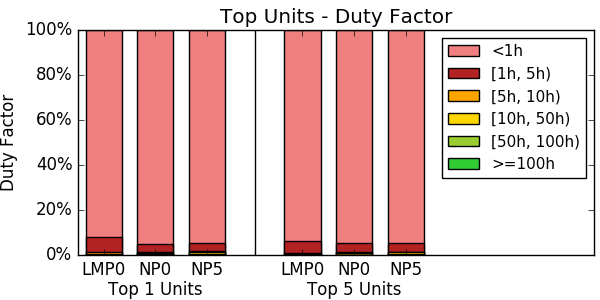}
\vspace{-0.1in}
\caption{Interval statistics by count for sites with highest duty factor.}
\vspace{-0.08in}
\label{figure:top-duty-interval-counts}
\end{figure}

\begin{figure}[htb]
\centering
\vspace{-0.1in}
\includegraphics[width=0.8\columnwidth]{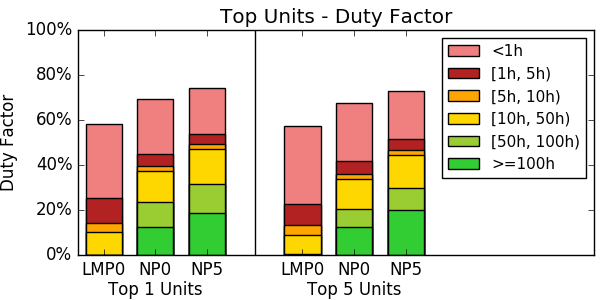}
\vspace{-0.1in}
\caption{Interval statistics by contribution to duty factor for sites with highest duty factor.}
\vspace{-0.08in}
\label{figure:top-duty-interval-contribution}
\end{figure}

Analyzing the interval statistics, Figure
\ref{figure:top-duty-interval-counts} shows that short intervals
dominate by count in both the top duty site, as well as the top 5 duty
factor sites.  In both cases, its important to note that the
preponderance of short intervals doesn't mean that there are no long
intervals.  In fact, most of the duty comes from intervals longer than
5 hours, and large fractions from intervals longer than 10 hours.  The
long intervals produce most of the duty factor (see Figure
\ref{figure:top-duty-interval-contribution}), accounting for nearly
50\% of the 70\% overall duty factor.
These results show that the top 5
duty factor sites are generally similar to the top site with similar
contribution to duty factor from long intervals.

\begin{figure}[htb]
\centering
\vspace{-0.1in}
\includegraphics[width=0.8\columnwidth]{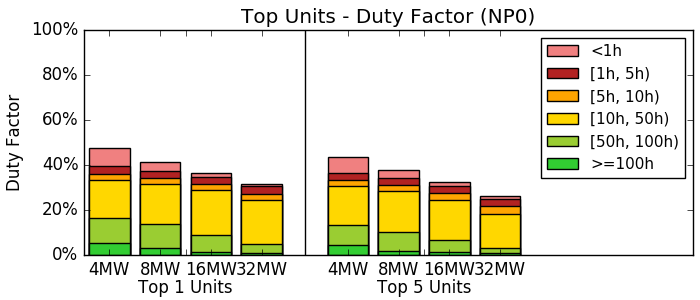}
\vspace{-0.1in}
\caption{Duty factor vs. power for sites with highest duty factor (np0).}
\vspace{-0.08in}
\label{fig:topduty_duty_vs_powerlevel_np0}
\end{figure}

We consider the duty factor, as a function of power level, characterizing the duty 
that could be supported with stranded power at a given power level (see Figure 
\ref{fig:topduty_duty_vs_powerlevel_np0}).  Loads of 4MW, 8MW, and 16MW can
be supported at duty factors of 40\%, 38\%, and 36\% respectively with intervals on average much
larger than a few hours.

\begin{figure}[htb]
\centering
\vspace{-0.1in}
\includegraphics[width=0.8\columnwidth]{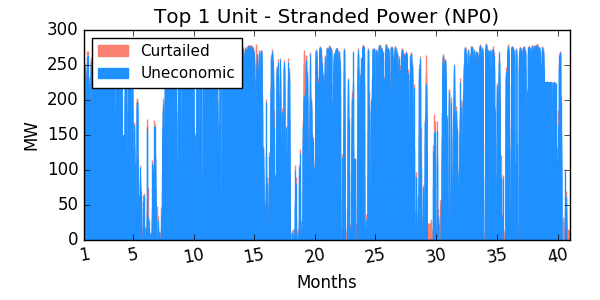}
\vspace{-0.1in}
\caption{Quantity vs. time for site with most stranded power.}
\vspace{-0.08in}
\label{fig:top1power_quantity_vs_time_np0}
\end{figure}

\begin{figure}[htb]
\centering
\vspace{-0.1in}
\includegraphics[width=0.8\columnwidth]{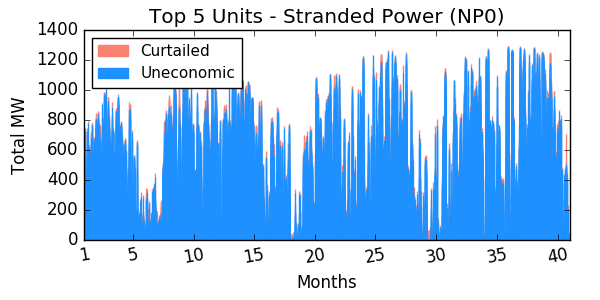}
\vspace{-0.1in}
\caption{Quantity vs. time for 5 sites with most stranded power.}
\vspace{-0.08in}
\label{fig:top5power_quantity_vs_time_np0}
\end{figure}

Considering sites with the most stranded power, we find over 250 MW power
(see Figure \ref{fig:top1power_quantity_vs_time_np0} and
\ref{fig:top5power_quantity_vs_time_np0}) and high variation.  On
these sites uneconomic power is far more than curtailed (Table
\ref{tab:topduty_sp}), and even considering the top 5 sites, power levels
as high as 100's of MW per site can be available.

\begin{figure}[htb]
\centering
\vspace{-0.1in}
\includegraphics[width=0.8\columnwidth]{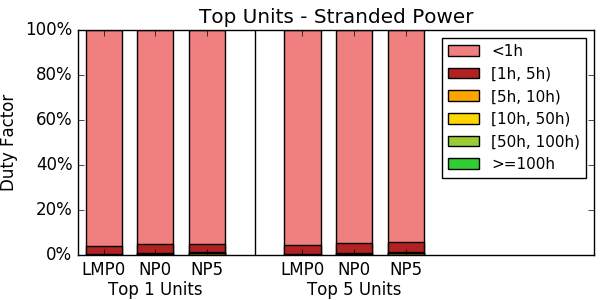}
\vspace{-0.1in}
\caption{Interval statistics by count for sites with most stranded power.}
\vspace{-0.08in}
\label{figure:top-power-interval-counts}
\end{figure}

\begin{figure}[htb]
\centering
\vspace{-0.1in}
\includegraphics[width=0.8\columnwidth]{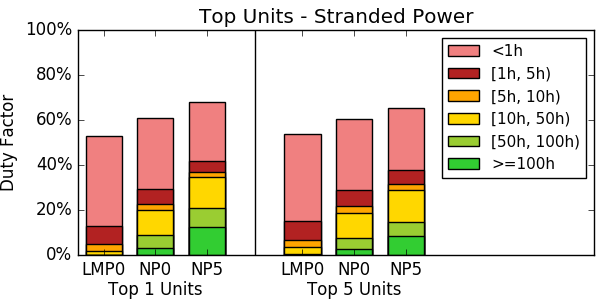}
\vspace{-0.1in}
\caption{Interval statistics by contribution to duty factor for sites with most stranded power.}
\vspace{-0.08in}
\label{figure:top-power-interval-contribution}
\end{figure}

Considering interval statistics for these sites with the most stranded power,
we find a similar story.  Figures
\ref{figure:top-power-interval-counts} and 
\ref{figure:top-power-interval-contribution} show 
that short intervals dominate by
count bue long intervals produce most
of the duty factor accounting for nearly
30\% of the 60\% overall duty factor

\begin{figure}[htb]
\centering
\vspace{-0.1in}
\includegraphics[width=0.95\columnwidth]{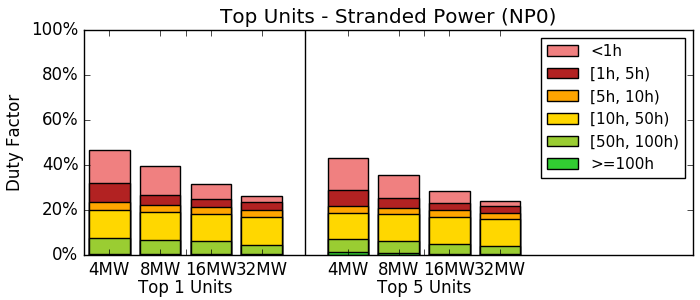}
\vspace{-0.1in}
\caption{Duty factor vs. power for Top sites with most stranded power (np0).}
\vspace{-0.08in}
\label{fig:toppower_duty_vs_powerlevel_np0}
\end{figure}

We consider the duty factor, as a function of power level, and here there is more
power available (see Figure 
\ref{fig:toppower_duty_vs_powerlevel_np0}).  Loads of 4MW, 8MW, 16MW, and 32 MW can
be supported at duty factors of 45\%, 40\%, 35\%, and 30\% respectively with intervals on average much
larger than a few hours.

\begin{table}[htb]
\vspace{-0.1in}
\caption{Statistics for Hourly NP0 Stranded Power (MW) For Top1 and Top5 Generators with Most Stranded Power (NP0)}
\label{tab:topduty_sp}
\vspace{-0.1in}
\centering
\begin{tabular}{l|r|r|r|r|r}
\hline
& Type & \#Hours & Mean & StdDev & Max \\
\hline \hline
Top1 & Uneconomic & 9,296 & 147.2 & 92.4 & 280.4 \\
\hline
Top1 & Curtailment & 26,790 & 5.6 & 8.5 & 215.9 \\
\hline
Top5 & Uneconomic & 15,569 & 330.2 & 300.0 & 1,285.7 \\
\hline
Top5 & Curtailment & 29,214 & 20.6 & 20.4 & 486.0 \\
\hline

\end{tabular}
\vspace{-0.08in}
\end{table}



\section{Enhancing Stranded Power}
\label{sec:enhanced}
\label{section:enhanced}

We consider enhancing the utility of stranded power by adding energy
storage \cite{CAISO-Storage-Pilots2024,TeslaPowerWall,CPUC13-storage}
to explore the incremental benefits and realizable improvements.  We
first consider global storage, then local storage, and finally
intelligently placed storage.  The major challenge with storage in the 
power grid is the vast scale of the power involved in the swings 
of variable generation.
We consider several scenarios of
grid-augmentation with storage that first bound the potential benefit and
then seek to characterize realistic benefits.

\begin{figure}[tbh]
\centering
\vspace{-0.1in}
\includegraphics[width=3.4in]{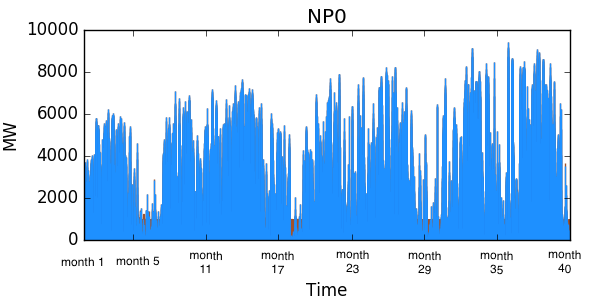}
\vspace{-0.1in}
\caption{Total Stranded power vs. time : with 8hr 1GW ideal global storage}
\vspace{-0.08in}
\label{fig:magic_global_storage_1GW8hr}
\end{figure}


The key functional parameters for energy storage include 1) {\it Rate} of
charge (MW) and 2) {\it Capacity} (MWH).
We consider a variety of capacities, charging rates, and connectivities.
And, because the desired use of power affects the best management of
the energy storage, our experiments involve a desired load.  
For {\it Load}, we use a constant level of power (e.g. 4MW,
8MW, 32MW, etc.), varying its
magnitude to assess stranded power utility.  Precisely, storage is managed as
in Algorithm \ref{algo:energy-storage-management}.

\begin{algorithm}
\caption{Energy Storage Management}
\label{algo:energy-storage-management}
\begin{algorithmic}
\IF{$strandedPower > load$}
\STATE ChargeAt max[$rate$,$strandedPower-load$]
\STATE Load is on
\ELSE
\IF{$storage + strandedPower \geq load$}
\STATE ChargeAt $(strandedPower - load)$
\STATE Load is on
\ELSE
\STATE ChargeAt $strandedPower$
\STATE Load is off
\ENDIF
\ENDIF
\end{algorithmic}
\end{algorithm}




\subsection{Global Energy Storage}

\begin{figure}[htb]
\centering
\vspace{-0.1in}
\includegraphics[width=3.4in]{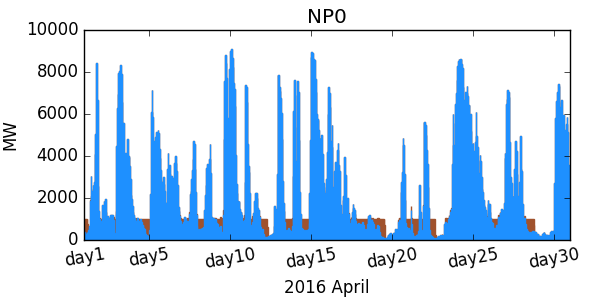}
\vspace{-0.1in}
\caption{Total power vs time (April 2016 - most Stranded Power month): With 8hr 1GW storage}
\vspace{-0.08in}
\label{fig:mostsp_month_8hr1GW}
\end{figure}

We first consider an idealized scenario,
global energy storage that can be directly connected to all parts
of the power grid.  This scenario ignores all location and transmission constraints,
so it provides an upper bound on the benefits of energy storage.
We add energy storage with 1GW rate and 8 hours capacity, comparable
to the aggregate scale of California's ``grid storage systems''
\cite{CPUC13-storage}, recently contracted (but configuration not
disclosed).  The results for MISO stranded power are shown in Figure
\ref{fig:magic_global_storage_1GW8hr} that displays the additional
power availability due to energy storage as brown (compare to Figure
\ref{fig:grid_quantity_vs_time_classify_np0}).  Because the impact is
small, we zoom in on the month with the most stranded power to
illustrate the detailed structure of how storage fills in when
stranded power is not available (Figure
\ref{fig:mostsp_month_8hr1GW}).

\begin{figure}[htb]
\centering
\vspace{-0.1in}
\includegraphics[width=0.75\columnwidth]{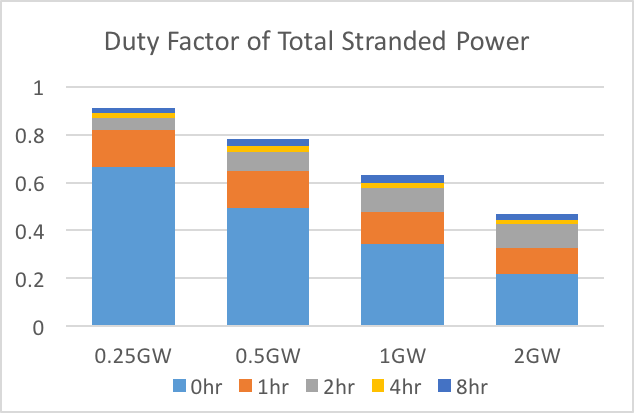}
\vspace{-0.1in}
\caption{Duty Factor of Ideal Global Storage, NP0}
\vspace{-0.08in}
\label{fig:magic_global_storage_df_np0}
\end{figure}

Adding storage does increase the duty factor of the MISO grid, but
despite ignoring transmission constraints (idealized global storage),
the incremental benefits fall off rapidly.  As shown in Figure
\ref{fig:magic_global_storage_df_np0}, adding 1 and 2 hours of storage
can improve the duty factor, at 0.25GW load, 1 hour of storage
increases duty factor to 82\% (a 15\% increase). But,benefits decrease
rapidly, with minimal benefits beyond 8 hours.  At higher power
levels, the achievable duty factor falls off significantly with 62\%
and 46\% achievable for 1 GW and 2 GW respectively.\footnote{Note that
  2 GWh at 2015 battery prices \cite{TeslaPowerWall} is over \$1.2B,
  or even 2nd generation variants (\$600M) or cheaper alternatives,
  comparable to the cost of approximately 200 MW installed wind
  turbines.}

\subsection{Uniformly Distributed Energy Storage}
\label{section:uniform-storage}


Next, we consider a more realistic scenario that distributes storage uniformly
to all wind generation sites.  This scenario assumes that transmission is
limited (or costly), and the storage is only used to
store the stranded power generated locally, and only provides power
locally.







\begin{figure}[htb]
\centering
\vspace{-0.1in}
\includegraphics[width=0.75\columnwidth]{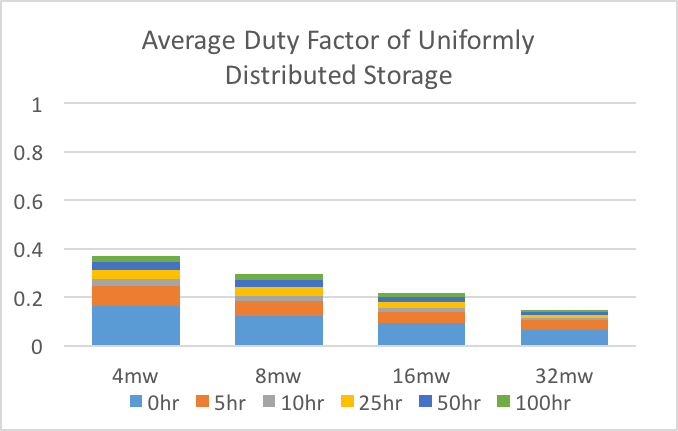}
\vspace{-0.1in}
\caption{Average of Site Duty Factors for Uniformly Distributed Energy Storage}
\vspace{-0.08in}
\label{fig:df_per_unit}
\end{figure}

We experiment with different amounts of energy storage, 5, 10, 25, 50,
and 100 hours for the load, but presume energy storage that can charge at the full rate of 
local generation (realistic for cell battery approaches, but perhaps not for lower cost options
such as pumped hydro or flow batteries).  
We compare to duty factor without storage as shown
in Figure \ref{fig:df_per_unit}.  The designated load affects the discharge 
time, so we vary the load from 4MW, 8MW,
16MW, to 32MW.  The corresponding storage capacity being added to the grid grows
as the product of duration and power level at each of the over 200 wind farm sites
as shown
in Table \ref{table:storage-capacity}.

\begin{table}
\caption{Storage Capacity added versus configuration (uniformly-distributed), MWh per site (grid = 200x)}
\label{table:storage-capacity}

\centering
\begin{tabular}{l|r|r|r|r|r}
\hline
           & 5 & 10 & 25 & 50 & 100 \\ \hline \hline
4 MW       & 20  & 40  & 100  & 200  & 400  \\ \hline
8 MW       & 40  & 80  & 200  & 400  & 800  \\ \hline
16 MW       & 80  & 160  & 400  & 800  & 1,600   \\ \hline
32 MW       & 160  & 320  & 800  & 1,600  & 3,200  \\ \hline
\end{tabular}
\end{table}

Our results show that 5 hours significantly increases the duty factor
at all power levels, but the benefits diminish quickly.  And that duty
factors of 20-45\% on average are achievable.  Note that these are the
average duty factors across the entire set of wind generation sites,
so we will see much higher duty factors for individual sites.  Note that these quantities
of storage go far beyond that economically viable.  For example, Tesla's Powerwall 2 is 
approximate \$300M per GWh, so even much cheaper storage would cost billions at this scale.

\begin{figure}[htb]
\centering
\vspace{-0.1in}
\includegraphics[width=3.4in]{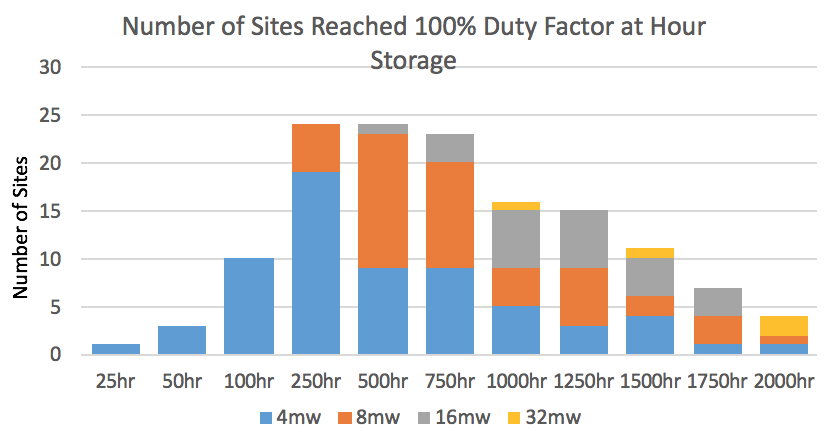}
\vspace{-0.1in}
\caption{\# Sites reaching 100\% Duty Factor at given Storage Capacity}
\vspace{-0.08in}
\label{fig:no_reached_100_1}
\end{figure}

Next we consider how much storage would be required to turn stranded
power, intermittent sites into reliable power sites (100\% duty).  That is, can we
easily turn stranded power into reliable continuous power?  As show in
Figure \ref{fig:no_reached_100_1}, it takes a large amount of storage
to make any sites reliable.  For 4 MW, a few sites can make it with
25-100 hours, but for any higher power levels, increases beyond 250
hours (10 days!) are required.  Referring to Table \ref{table:storage-capacity},
50 hours of storage for 4-32MW, ranges from 200MWh to 1,600 MWh per site, or approximately \$50M to \$400M, comparable to the wind turbine costs.



\subsection{Intelligently Placed Storage}
\label{section:uniform-storage}

Uniform distribution of storage 
places it inefficiently,
adding it where little stranded power occurs, and thus deriving less
benefit.  We consider intelligent placement, picking to two
types of promising sites -- those where the maximum quantity of stranded
power is being generated, and those where the maximum duty factor of stranded power
occurs.

\begin{figure}[htb]
\centering
\vspace{-0.1in}
\includegraphics[width=0.75\columnwidth]{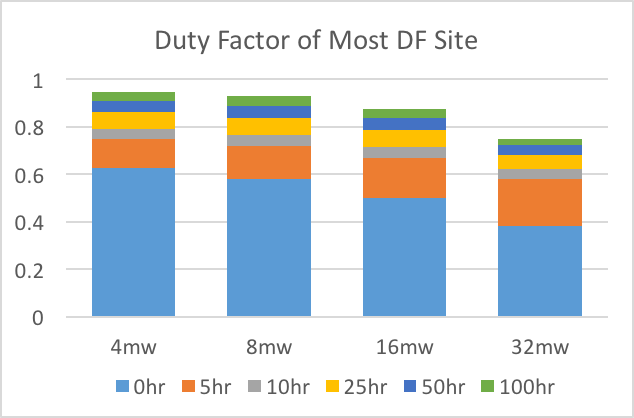}
\vspace{-0.1in}
\caption{Storage benefits Duty Factor of Highest Duty Factor Site(single site),NP0}
\vspace{-0.08in}
\label{fig:mostdf_storage_df_np0}
\end{figure}

\begin{figure}[htb]
\centering
\vspace{-0.1in}
\includegraphics[width=0.75\columnwidth]{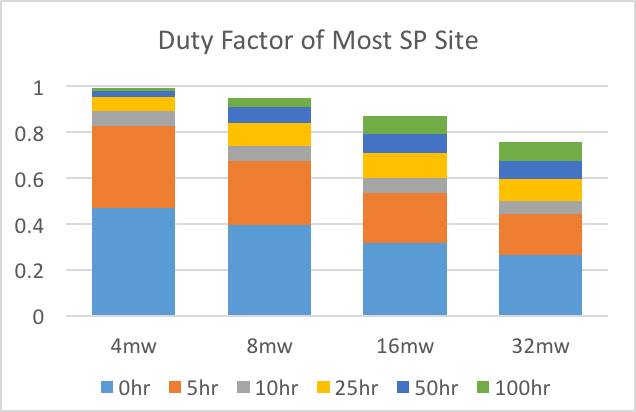}
\vspace{-0.1in}
\caption{Storage benefits Duty Factor of Most Stranded Power Site(single site), NP0}
\vspace{-0.08in}
\label{fig:mostsp_storage_df_np0}
\end{figure}

Figures \ref{fig:mostdf_storage_df_np0}
and \ref{fig:mostsp_storage_df_np0} show that storage placed
at the top duty factor and top stranded power
sites gives much greater benefits.  From a much higher base, adding 5 hours of storage can raise
duty factors to as much as 75-85\%.  And even at higher power levels
of 32 MW, storage is highly productive, and can lift duty factors
to over 75\%.  These benefits and absolute duty factors achieved are more than
2x better for uniformly-distributed storage (Figure \ref{fig:df_per_unit}).
Clearly, investment in expensive energy storage will be yield much greater benefits
at these few top sites, compared to uniformly distributed storage.

\begin{figure}[htb]
\centering
\vspace{-0.1in}
\includegraphics[width=0.75\columnwidth]{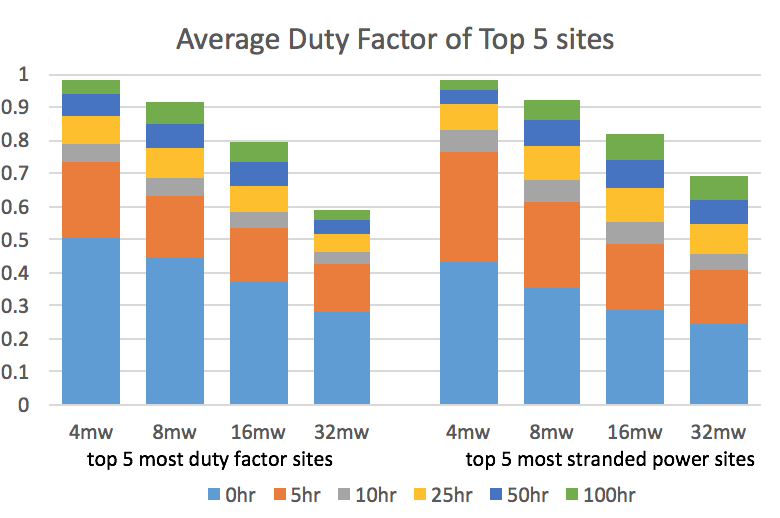}
\vspace{-0.1in}
\caption{Average Duty Factor with energy storage, Top 5 Sites, NP0}
\vspace{-0.08in}
\label{fig:top5sp_storage_df_np0}
\end{figure}

To see how robust these differences are, we consider the top 5 in each of the duty factor 
and stranded power (Figure \ref{fig:top5sp_storage_df_np0})
categories.  Again these results show that much greater storage
benefits can be achieved at the top duty factor and top stranded power
sites.   With 5 hours, 
duty factors reach as much as 75\%, and at higher power levels
up to 32 MW, storage's benefits are robust across a collection of sites.

\subsection{Incremental Benefit of Storage Capacity}



\begin{figure}[htb]
\centering
\vspace{-0.1in}
\includegraphics[width=3.4in]{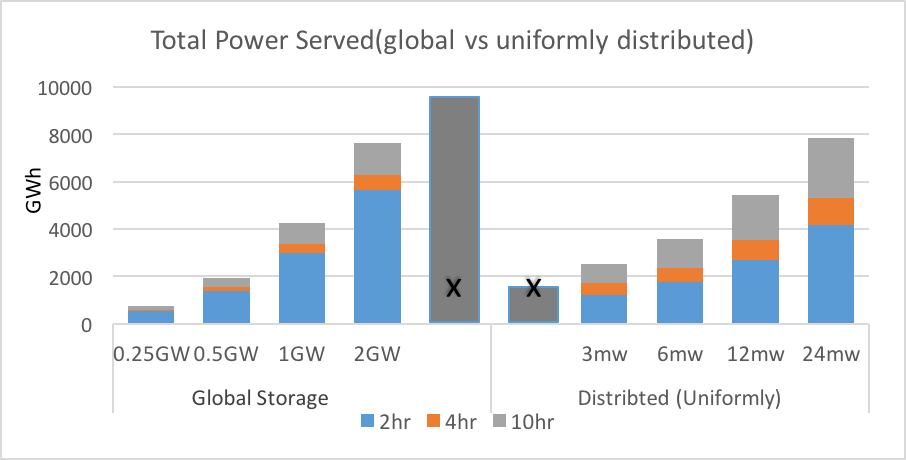}
\vspace{-0.1in}
\caption{Total Power Served by energy storage: Global vs. Uniformly distributed}
\vspace{-0.08in}
\label{fig:total_power_served_global}
\end{figure}

\begin{figure}[htb]
\centering
\vspace{-0.1in}
\includegraphics[width=3.4in]{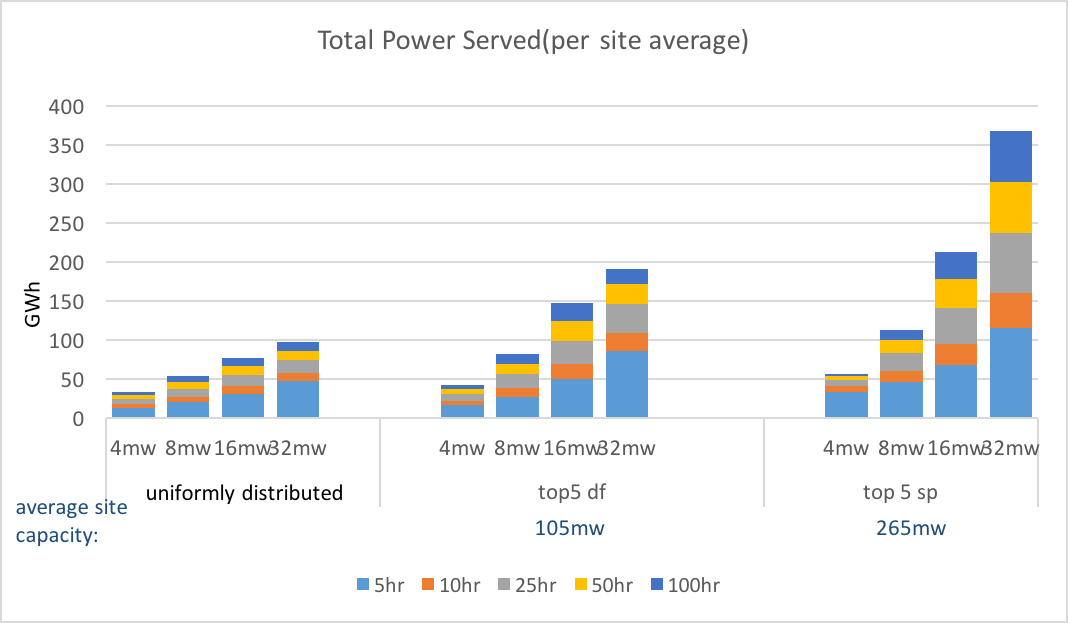}
\vspace{-0.1in}
\caption{Total Power Served(Site Average) of Per Site Placed Storage}
\vspace{-0.08in}
\label{fig:total_power_served_per_site}
\end{figure}

We evaluate the effectiveness of storage placement by
the total power it serves (captures).  In
Figure \ref{fig:total_power_served_global}, we compare global and
uniformly distributed.  Interestingly, the comparison shows that
global storage is only slightly better than uniformly distributed.
Comparing two intelligent distribution schemes (see
Figure \ref{fig:total_power_served_per_site}) show clearly that energy storage
deployed at the top 5 sites can be as much as 3.7 times more
productive.



\begin{figure}[htb]
\centering
\vspace{-0.1in}
\includegraphics[width=3.4in]{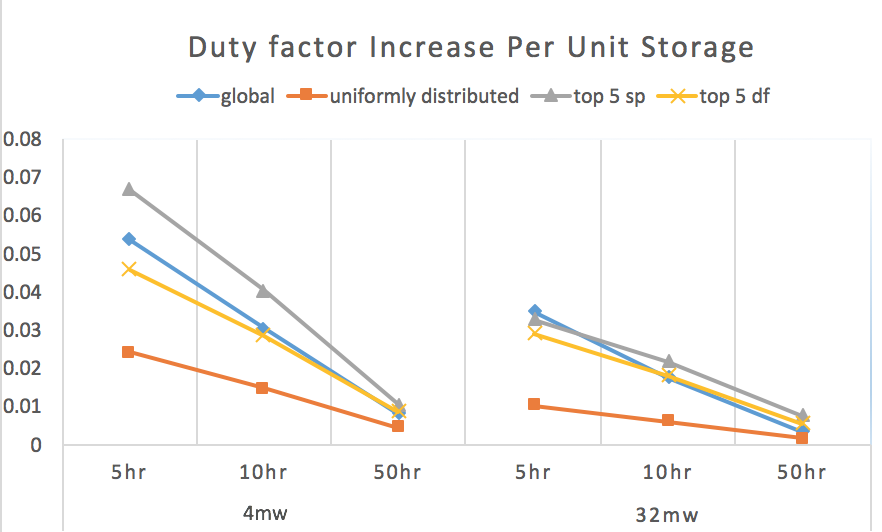}
\vspace{-0.1in}
\caption{Duty Factor Increase per Unit/Hour Storage}
\vspace{-0.08in}
\label{fig:dfincrease_per_unit_side}
\end{figure}

To get directly at cost-benefit, we plot the duty factor benefit per
unit storage, for a variety of power levels and storage capacities
(see Figure \ref{fig:dfincrease_per_unit_side}).  These results show a
rapid downward trend as storage capacity increases, 
going from 5 to 50 hours, the benefit per unit decreases by 4-10 fold.
This suggests that even if small quantities of storage become
economically viable, large
quantities may not.



\section{Discussion and Related Work}
\label{sec:related_work}
\label{section:related}

Stranded power has been documented as a large,
growing untapped resource.  Published reports 
document that the Midcontinent Independent System
Operator (MISO) power grid
curtailed 2.2 terawatt-hours (TWh) of power (see Figure
\ref{figure:stranded_power_lmp0}) and bought 5.5 TWh at negative price for a total of
7.7 TWh of stranded power from wind resources
\cite{CurtailmentUS14,Bird2013,YangChien15}.  And in China, curtailed
wind power has grown to 34 TWh in 2015 \cite{GWEC-Annual16},
suggesting stranded power of 60-100 TWh.  Around the world, as
renewable generation fraction increases due to rising RPS standards,
stranded power is projected to increase significantly
\cite{Bird2013,E3report,KYZC2016}.
All of these efforts report aggregated curtailment over entire dispatch
regions, and often at a coarse temporal granularity (months or years).  The
reason is that these studies are concerned with effectiveness of incorporating
renewable-generators power with the goal of reducing carbon emissions,
and grid generator economic viability.  To our knowledge, this study is the first that looks at
fine-grained curtailed power availability, and add to that a fine-grained 
temporal analysis of uneconomic power -- together stranded power.

Numerous studies explore challenges in renewable integration,
assessing concerns of grid stability, ability to achieve ``merit
order'', and also the dynamics of markets \cite{Bird2013,Solar-Challenges16,E3report}.  
These studies generally point out the daunting challenges to grid operations as RPS
standards continue to rise, and suggest that the stranded power we
have studied is a significant and persistent phenomena.  In fact
several recent studies suggest that stranded power likely to grow
rapidly in both wind-heavy \cite{KYZC2016} and solar-heavy
\cite{Storage-California-50RPS-2016,E3report,Solar-Challenges16} renewable power grids.  Such results
suggest that the study of stranded power is of increasing importance.

In addition to broad demand-response studies,
and specific load shifting and demand-response \cite{battery,buildings,alcoa},
several studies explore data-center demand-response (DCDR)
\cite{Wierman-DCDR14}; 
in other work,
we have proposed a new model
where 
data centers can be ``dispatchable loads'' based on new computer science
approaches to create flexible computing loads \cite{chien2015zero,yang2016zccloud,Chien-Dispatch-Load15,TR-ZCCloud2016}. These
data centers have the dynamic range for load (10's of MW) that can exploit stranded power
at the scales we have characterized.

Progress in energy storage is promising \cite{Storage-Roadmap2015},
but the low price of power makes its large-scale deployment for
time-shifting challenging \cite{Storage-California-50RPS-2016} and
some studies suggest on energy return-on-investment (EROI) criteria,
it may never make sense to store large quantities of wind power
\cite{EROI2013}.  Adding energy storage to distributed (residential)
solar has recently become popular \cite{TeslaPowerWall}, but
grid-scale storage for time-shifting, rather than small quantities for
peak shaving and regulation, faces significant economic challenges.
Recent years has seen the deployment of experimental scale battery
storage, primarily for regulation
\cite{CAISO-Storage-Pilots2024,CPUC13-storage} and the ancillary
services market.  It is worth noting that the largest of these, in
CAISO, will deploy 1.3 GW of storage by 2024, a small fraction of the
quantity of storage considered in Section \ref{sec:enhanced}, and
our results (recall Figure \ref{fig:no_reached_100_1}) shows diminishing
returns for each increment of storage.

\section{Summary and Future Work (0.25 pages)}
\label{sec:summary}
\label{section:summary}

We have presented a detailed temporal and spatial study of stranded
power (curtailed and uneconomic power) arising in a modern power grid.
Our results show the large magnitude (terawatt-hours) and nature of
this phenomenon (intermittent, with long intervals) in a power grid
with wind-based renewable generation.  The magnitude of available
stranded power in long intervals (multiple days) suggests that
exploitation may be possible.  The MISO grid we studied has a low RPS,
only 10\% wind, suggesting that higher RPS grids may have even more attractive
stranded power resources.

Study of enhancing usability of stranded power with storage suggests
that while small quantities of storage can signficantly increase duty
factors, and by choosing sites carefully, the effectiveness can be even
greater.  However, approaching 100\% (reliable grid power) would require extremely
large quantities of storage.
Promising directions for future research include study of a high-RPS solar
grid such as CAISO and higher-RPS grids.









\section*{Acknowledgment}
This work was supported by in part the National Science Foundation
under Award CNS-1405959, and the Office ofAdvanced Scientific Computing
Research, Office of Science,U.S.  Department of Energy, under Award
DE-SC0008603and Contract DE-AC02-06CH11357.  We also
gratefully acknowledge generous support Keysight and
the Seymour Goodman Foundation.

\ifCLASSOPTIONcaptionsoff
  \newpage
\fi



\bibliographystyle{IEEEtran}


\def\UrlBreaks{\do\/\do-\do\.\do:}
\bibliography{zccloud,zccloud2,zccloud3,zccloud4}
\end{document}